\documentclass[aps,prb,twocolumn,superscriptaddress]{revtex4}
\usepackage{dcolumn} 
\usepackage{graphicx}
\usepackage{bm}      
\usepackage{xcolor}
\usepackage{amsmath,amssymb}
\usepackage{hyperref}
\usepackage{multirow}
\usepackage{epstopdf}
\usepackage{latexsym}
\usepackage{subfigure}
\usepackage{wasysym} 
\usepackage{times}
\usepackage{gensymb}
\usepackage{textcomp}

\def \bfso{Ba$_{2}$FeSi$_{2}$O$_{7}$}

\usepackage[normalem]{ulem}

\begin{document}

\title{Field-induced spin level crossings within a quasi-{\it XY} antiferromagnetic state in \bfso{}}
\author{Minseong Lee}
\email{ml10k@lanl.gov}
\affiliation{National High Magnetic Field Laboratory, Los Alamos National Laboratory, Los Alamos, NM 87545, USA.}
\author{Rico Schoenemann}
\affiliation{National High Magnetic Field Laboratory, Los Alamos National Laboratory, Los Alamos, NM 87545, USA.}
\author{Hao Zhang}
\affiliation{Materials Science and Technology Division, Oak Ridge National Laboratory, Oak Ridge, TN, USA.}
\affiliation{Department of Physics and Astronomy, University of Tennessee, Knoxville, TN, USA.}
\author{David Dahlbom}
\affiliation{Department of Physics and Astronomy, University of Tennessee, Knoxville, TN, USA.}
\author{Tae-Hwan Jang}
\affiliation{MPPHC-CPM, Max Planck POSTECH/Korea Research Initiative, Pohang, Republic of Korea.}
\author{Seung-Hwan Do}
\affiliation{Materials Science and Technology Division, Oak Ridge National Laboratory, Oak Ridge, Tennessee 37831, USA.}
\author{Andrew D. Christianson}
\affiliation{Materials Science and Technology Division, Oak Ridge National Laboratory, Oak Ridge, Tennessee 37831, USA.}
\author{Sang-Wook Cheong}
\affiliation{MPPHC-CPM, Max Planck POSTECH/Korea Research Initiative, Pohang, Republic of Korea.}
\affiliation{Rutgers Center for Emergent Materials and Department of Physics and Astronomy, Rutgers University, Piscataway, NJ, USA.}
\author{Jae-Hoon Park}
\affiliation{MPPHC-CPM, Max Planck POSTECH/Korea Research Initiative, Pohang, Republic of Korea.}
\affiliation{Department of Physics, Pohang University of Science and Technology, Pohang, Republic of Korea.}
\author{Eric Brosha}
\affiliation{Materials Synthesis and Integration, Los Alamos National Laboratory, Los Alamos, NM 87545, USA.}
\author{Marcelo Jaime}
\affiliation{National High Magnetic Field Laboratory, Los Alamos National Laboratory, Los Alamos, NM 87545, USA.}
\author{Kipton Barros}
\affiliation{Theory Division, Los Alamos National Laboratory, Los Alamos, NM 87545, USA.}
\author{Cristian D. Batista}
\email{cbatist2@utk.edu}
\affiliation{Department of Physics and Astronomy, University of Tennessee, Knoxville, TN, USA.}
\author{Vivien S. Zapf}
\email{vzapf@lanl.gov}
\affiliation{National High Magnetic Field Laboratory, Los Alamos National Laboratory, Los Alamos, NM 87545, USA.}
\date{\today}

\begin{abstract}
We present a high-field study of the strongly anisotropic easy-plane square lattice $S$ = 2 quantum magnet Ba$_{2}$FeSi$_{2}$O$_{7}$. This compound is a rare high-spin antiferromagnetic system with very strong easy-plane anisotropy, such that the interplay between spin level crossings and antiferromagnetic order can be studied.
We observe a magnetic field-induced spin level crossing occurring within an ordered state. This spin level crossing appears to preserve the magnetic symmetry while  producing a non-monotonic dependence the order parameter magnitude.
 The resulting temperature-magnetic field phase diagram exhibits two dome-shaped regions of magnetic order overlapping around 30 T. The ground state of the  lower-field dome is predominantly a linear combination of $| S^{z} = 0 \rangle$ and $ |S^{z} = 1 \rangle$ states, while the ground state of the higher-field dome can be approximated by a linear combination of $| S^{z} = 1 \rangle $ and $ | S^{z} = 2\rangle$ states. At 30 T, where the spin levels cross, the magnetization exhibits a slanted plateau, {\color {black}the magnetocaloric effect shows a broad hump, and the electric polarization shows a weak slope change}. We determined the detailed magnetic phase boundaries and the spin level crossings using measurements of magnetization, electric polarization, and the magnetocaloric effect in pulsed magnetic fields to 60 T. We calculate these properties using a mean field theory based on direct products of SU(5) coherent states and find good agreement. Finally, we measure and calculate the magnetically-induced electric polarization that reflects magnetic ordering and spin level crossings. This multiferroic behavior provides another avenue for detecting phase boundaries and symmetry changes.
\end{abstract}

\maketitle
\section{INTRODUCTION}

\begin{figure*}[tbp]
	\linespread{1}
	\par
	\begin{center}
		\includegraphics[width=0.9\textwidth]{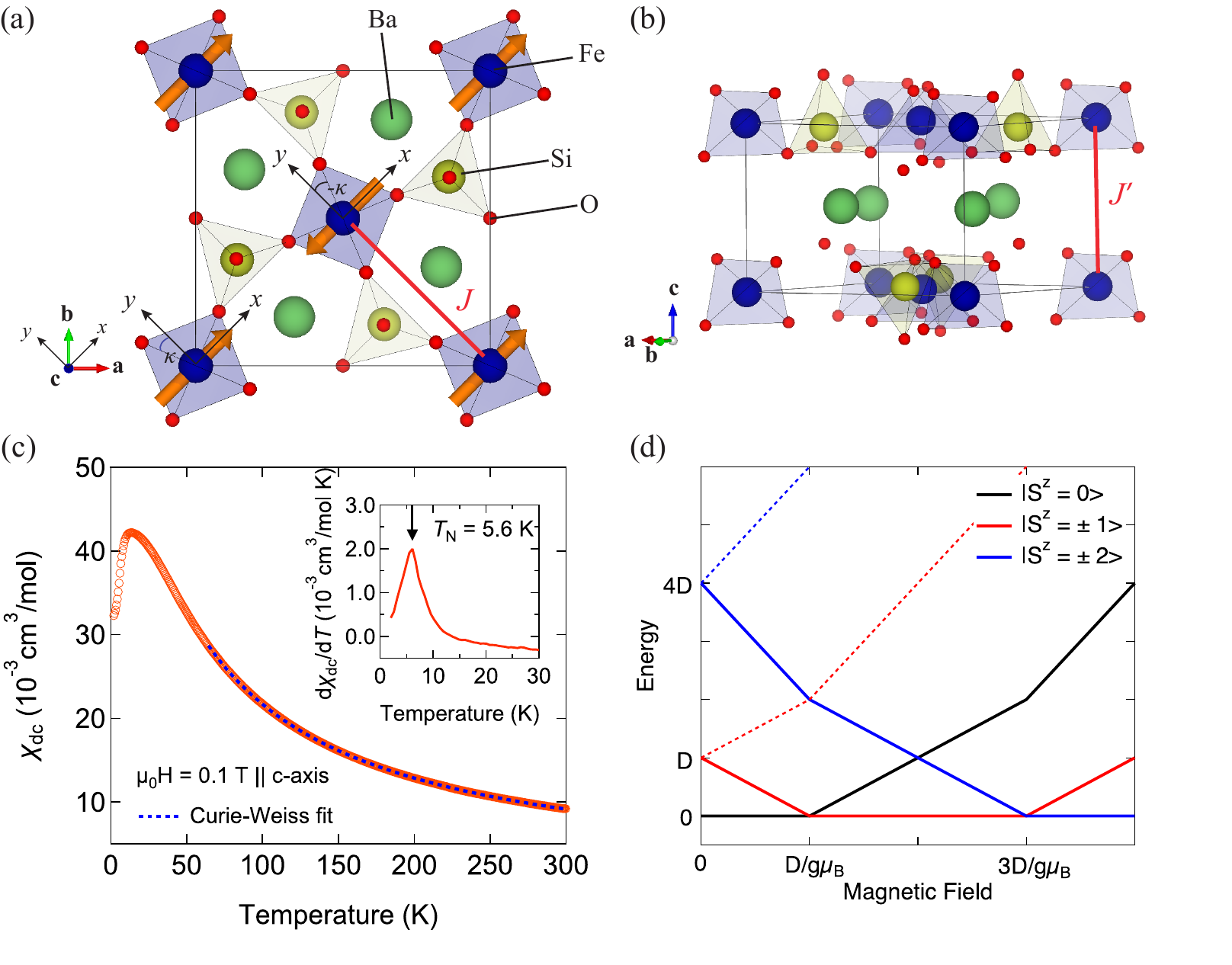}
	\end{center}
	\par
	\caption{\label{fig_crystalstructure} {\bf Crystal structure and DC magnetic susceptibility of \bfso{}}. (a) Tetrahedra of FeO$_{4}$ interconnected by nonmagnetic SiO$_{4}$ form a square lattice in $ab$-plane. The nearest super-exchange interaction is denoted as $J$. (b) Each $ab$-plane is separated by large Ba$^{2+}$ ions and the interlayer coupling is denoted as $J'$, which is much weaker than $J$. (c) The temperature dependence of the {\color {black} DC} magnetic susceptibility. The symbols are experimental data and the dashed line represents the Curie-Weiss fit. Inset: temperature derivative of the {\color {black} DC} magnetic susceptibility, in which the peak signals the antiferromagnetic long-range ordering. The spin structure is presented in (a) with orange arrows. (d) The spin energy levels as a function of magnetic field, neglecting magnetic exchange broadening. 
	}
\end{figure*}

Antiferromagnets with strong easy-plane magnetic anisotropy 
are a rich source of quantum magnetism and quantum phase transitions \cite{intro_1,intro0,intro01}. In the limit of strong easy-plane single-ion anisotropy,  the Hamiltonian contains a term $\mathcal{H}_D = \sum_{i}D(S^z_i)^2$ with $D > 0$ that is dominant over the other interactions. This term splits the $2{S}+1$ spin levels into an $S^{z} =  0$ ground state and excited states with $S^{z} =  \pm 1,\pm 2, \cdots$, $\pm S$,  separated by energy gaps of $1^2D$, $2^2D$, $\cdots$, $S^2D$ for integer spin. The $z$ direction coincides with the uniaixal anisotropy direction created by $\mathcal{H}_D$.  Since the Zeeman term generated by an external magnetic field parallel to the $z$-direction commutes with $\mathcal{H}_D$, the eigenstates of the single-ion Hamiltonian are preserved and the energy gaps evolve linearly with the field. The energy gaps between positive $S^{z}$ states and the ground state are suppressed by the field, resulting in successive spin level crossings.  Small exchange interactions between neighboring ions $J < D$ disperse and broaden these levels, producing a mean field ground state that is direct product of  linear combinations of different $|S^{z} \rangle$ states. 
The expectation value of the in-plane component of the local magnetization is finite in this wave function, leading to quasi-{\it XY} antiferromagnetic long-range ordering occurring in dome-shaped regions of the temperature-magnetic ($T-H$) phase diagram \cite{intro4,Samulon09,Jaime04}. The order parameter (planar staggered magnetization) can be described as a complex number, whose amplitude is the magnitude of the planar magnetization, while the phase determines the direction of the spins. The pioneering work of Matsubara and Matsuda showed that interacting spin-$1/2$ systems can be mapped into a interacting bosonic gas with hard-core repulsion \cite{intro1}. This mapping can be generalized for higher spin systems \cite{intro2,intro3}, implying that field-induced quantum phase transitions of magnets with uniaxial symmetry belong to the Bose-Einstein condensation (BEC) universality class. In the presence of disorder or competing interactions, Bose glass \cite{boseglass01,boseglass02,boseglass03,boseglass04} or even super-solidity \cite{supersolid01,supersolid02,supersolid03,supersolid04} can also occur. It is important to note, however, that the Hamiltonians of real spin systems  always include  terms that break the uniaxial symmetry, but these terms are small for many compounds and the BEC description becomes an excellent approximation, particularly of the quantum phase transitions and their excitations.\cite{intro0}

Most work in this field has been done on ${S} = 1/2$ dimers \cite{TlCuCl01,Jaime04} or monoatomic ${S} = 1$ systems \cite{intro0,DTN02}. Higher spin systems (${S} > 1$) with strong easy-plane anisotropy $D \gg J$ are rare. However, when they occur, the energy of more than one high-spin states can be lowered by magnetic field to become the ground state in high fields, creating multiple domes of antiferromagnetism in $T-H$ space. For example, Ba$_3$Mn$_2$O$_8$ forms exchange-coupled dimers of Mn$^{5+}({S} = 1)$ spins \cite{Samulon09}. In applied magnetic fields, two dome-shaped regions of antiferromagnetic order can be observed between 10 and 25 T and between 35 and 41 T.
However, to our knowledge there is no example of monoatomic ${S} = 2$ systems with strong easy-plane anisotropy that show this behavior. When the energy scales for single-ion anisotropy and magnetic exchange are not well separated, the low and high-field domes overlap leading to a single {\color {black} double-peaked region of antiferromagnetism} with {\color {black}the unusual feature that the $S_z$ state evolves within the dome.} Here we investigate this phenomenon in the $\bf{S} = 2$ easy-plane antiferromagnet \bfso{}. The compound belongs to the Melilite family  $A_{2}$$M$$B_{2}$O$_{7}$ ($A$ = Ca, Sr, Ba, $M$ = divalent 3\textit{d} transition metals, $B$ = Si, Ge) and forms in the P$\bar{4}2_{1}$m tetragonal structure. 

The relatively strong spin-orbit coupling of Fe$^{2+}$ ($\sim$ 20 meV) ions combined with the crystal field from the largely compressed FeO$_{4}$ tetrahedra along $c$-axis is identified as the source of an unusually large easy-plane anisotropy from a THz spectroscopic study \cite{SFSO}. T.-H. Jang {\it et al.} recently reported the growth of high-quality single crystals of \bfso{} and thermodynamic investigations show very strong easy-plane single-ion magnetic anisotropy and the indication of well-separated spin levels from the specific heat measurement \cite{THJang_PRB}.  Neutron diffraction measurement at zero magnetic field finds antiferromagnetism with spins aligned along the diagonal direction of the $a$- and $b$- axes as shown in \autoref{fig_crystalstructure} (a). At zero field, S.-H. Do {\it et al.}, observed the emergence, strong decay, and renormalization of a longitudinal magnon mode in \bfso {} since the compound is close to a quantum critical point due to the large ratio of easy-plane anisotropy to the Heisenberg exchange interaction ($D/J$) \cite{intro7}. These experimental findings motivated us to investigate the interplay of magnetic order and spin level crossings at high magnetic fields. 

Similar to the case of Ba$_3$Mn$_2$O$_8$, \bfso{} 
exhibits two field-induced domes associated with local mean field states that are  mostly linear combinations of the $| S^{z} = 0\rangle$ and $ | S^{z} = 1\rangle$ states for the low-field dome  and the $| S^{z} = 1\rangle $ and $ | S^{z} = 2\rangle$  states for the high-field dome. However, in contrast to the Mn system, for \bfso{} there is finite overlap between both domes corresponding to a  field-induced crossover region between both {\it XY} AFM phases. This means a spin level crossing occurs within the AFM phase. Also, as shown in \autoref{fig_crystalstructure}~(c), the low-field dome  extends down to zero field with a N{\'e}el temperature $T_{\text{N}} = 5.6$ K. {\color {black} We note that the spin levels are broadened by exchange and thus can also be thought of as bands, which cross each other over finite regions of magnetic field.} We also measure and calculate the multiferroic behavior in the {\it p-d} orbital hybridization model \cite{Kim_multiferroic}.  Finally, we build a spin Hamiltonian and solve it using a mean field theory based on SU(5) coherent states. This is the starting point of a generalized spin-wave theory that captures the high-energy modes associated with $\left|S^z=\pm 2\right>$ states~ \cite{Muniz,HZhang_GMFT,Dahlbom22,Dahlbom22b}. The SU(5) coherent states are strictly necessary to study the evolution of different properties as a function of magnetic field because the $\left|S^z=2\right>$ state becomes a low-energy excitation and eventually the ground state for high enough field parallel to the $c$-axis.  
We demonstrate that the resulting mean field wave functions describe the magnetization and electric polarization data over the whole range of $T$ and $H$ remarkably well.

\section{EXPERIMENTAL METHODS}
We first prepared a polycrystalline sample of \bfso{} as a precursor using the solid-state reaction. A stoichiometric mixture of BaCO$_{3}$, Fe$_{2}$O$_{3}$, and SiO$_{2}$ were thoroughly ground, pelletized, and heated at 1050$^{\circ}$C with intermediate sintering. X-ray and neutron powder diffraction measurements on the polycrystalline samples identified dominant phase of \bfso{} (96.5\%) \cite{THJang_PRB}. The polycrystalline samples were prepared as feed rods, and a single crystal of \bfso{} was grown using floating zone melting method under a reducing gas atmosphere. The high quality of the single crystal has been also confirmed using X-ray diffraction \cite{THJang_PRB}.

The magnetization vs. magnetic field in millisecond pulsed magnets up to 60 T was measured at the pulsed field facility of Los Alamos National Laboratory. A single crystal with $H\parallel c$ was measured in an ampoule secured with Apiezon N grease. The sample was placed in a radially counterwound copper coil that subtracts the background signal of the spatially-uniform pulsed magnetic field. \cite{Detwiler00} The sample's own magnetic moment is non-uniform on the scale of the coil. The change in this magnetic moment with pulsed magnetic field is recorded and integrated in time. The drift in the compensation of the coil due to thermal contraction is compensated for by subtracting a portion of the signal from an extra loop of wire, and finally an additional background is subtracted by measuring with the sample in and out of the coil with an in-situ extraction rod. Helium-3 or Helium-4 were used to thermalize the samples and a Cernox thermometer from Lakeshore recorded $T$ right before each field pulse. The pulsed-field magnetization is calibrated using {\color {black} DC}-magnetization measurements with a Vibrating Sample Magnetometer in a 14 T PPMS by Quantum Design. 

In order to measure the magnetocaloric effect (temperature changes vs magnetic field) in pulsed magnetic field, an AuGe thin film thermometer was directly deposited on the surface of the sample by RF magnetron sputtering at 40 mTorr pressure of ultra-high purity Ar gas for 60 minutes with 100 W power. We also prepared a reference AuGe film on a glass plate. The resistance of the sample and reference thermometers were calibrated against a Lakeshore Cernox thermometer at zero field. We recorded the resistance of AuGe films as a function magnetic field and the background magnetoresistance was subtracted using the reference thermometer. 

For the electric polarization measurement in pulsed magnetic fields, two single crystals shaped as thin plates were measured in a parallel-plate capacitance geometry with typical dimensions of 3 $\times$ 3 $\times$ 0.5 mm$^{3}$. Measurements were made for $P \parallel \rm{c}$ and $P \perp \rm{c}$ with $H\parallel c$. Silver paint was applied to parallel surfaces of the sample to form capacitor plates and one plate was grounded and the other virtually grounded through a Stanford Research 570 current amplifier, such that charge is pulled out of ground to compensate the electric polarization of the sample. The electric polarization was obtained by the standard method \cite{Zapf10} of recording the current between the sample's capacitor plate and ground with the SR570 due to the sample's changing electric polarization with field. This current can be integrated in time to obtain the electric polarization.

All three of these measurements benefited in some aspects from the speed of pulsed fields - magnetization and electric polarization signals are proportional to $dH/dt$, whereas the noise scales only like the square root of the data collection time. Meanwhile, the magnetocaloric effect is pushed closer to the adiabatic regime by the fast speed of the pulses.

\section{RESULTS}
\label{results}
\autoref{fig_crystalstructure} shows the crystal structure and {\color {black} DC} magnetic susceptibility of \bfso{}. \bfso{} crystallizes in a piezoelectric (noncentrosymmetric, nonpolar) tetragonal structure that belongs to the P${\bar 4}2_{1}$m (113) space group. This means electric polarization can be induced by certain uniaxial lattice strains that creates a polar axis.
\autoref{fig_crystalstructure} (a) shows that the magnetic ion Fe$^{2+}$ (${S} = 2$) is tetrahedrally coordinated by oxygen atoms,  which are interconnected via nonmagnetic tetrahedra of SiO$_{4}$ to form a  square lattice layer. The antiferromagnetic nearest-neighbor exchange interaction ($J$) within the plane is mediated by two O atoms and one Si atom. Large nonmagnetic Ba$^{2+}$ ions effectively disconnect the square lattice planes, as shown in \autoref{fig_crystalstructure} (b). Thus, the interlayer coupling ($J'$) is ten times weaker than $J$ (\autoref{parameter_table}).
 These crystal structures have been resolved using X-ray diffraction and neutron diffraction measurements in the previous study \cite{THJang_PRB}. {\color {black} DC} magnetic susceptibility of \bfso{} between 2 K and 300 K with the magnetic field of 0.1 T along $c$-axis is presented in \autoref{fig_crystalstructure} (c).  We fit a Curie-Weiss law (see S.I.\cite{SI}) $\chi (T) = C/(T-\Theta_{\text{CW}}) + \chi_{0}$ to the data between 100 K and 300 K where $C$ is Curie constant and $\Theta_{\text{CW}}$ is the Curie-Weiss temperature. From the fitting, 
we obtain $\mu_{\text{eff}} = 5.03 \mu_{\text{B}}$, which is comparable to the spin-only value of ${S}= 2$ ($\mu_{\text{eff}} = 4.90 \mu_{\text{B}}$ for $g$ = 2), and $\Theta_{\text{CW}}$ = $-$45.7 K, which reflects both the antiferromagnetic exchange and the easy-plane anisotropy. $\chi_{0}$ is the diamagnetic contribution, which is three orders of magnitude smaller than the sample signal.
The magnetic susceptibility displays a broad hump around 20~K, which we attribute to the level splitting caused by the single-ion anisotropy term $\mathcal{H}_D$ \cite{THJang_PRB}. Below 20~K, the susceptibility plummets quickly at the onset of three dimensional antiferromagnetic long-range ordering.
As shown in the inset of \autoref{fig_crystalstructure}~(c), we found a N\'eel temperature ($T_{\text{N}}$) of 5.6~K from the peak in the first derivative of {\color {black} DC} magnetic susceptibility.

\begin{figure}[tbp]
	\linespread{1}
	\par
	\begin{center}
		\includegraphics[width=0.5\textwidth]{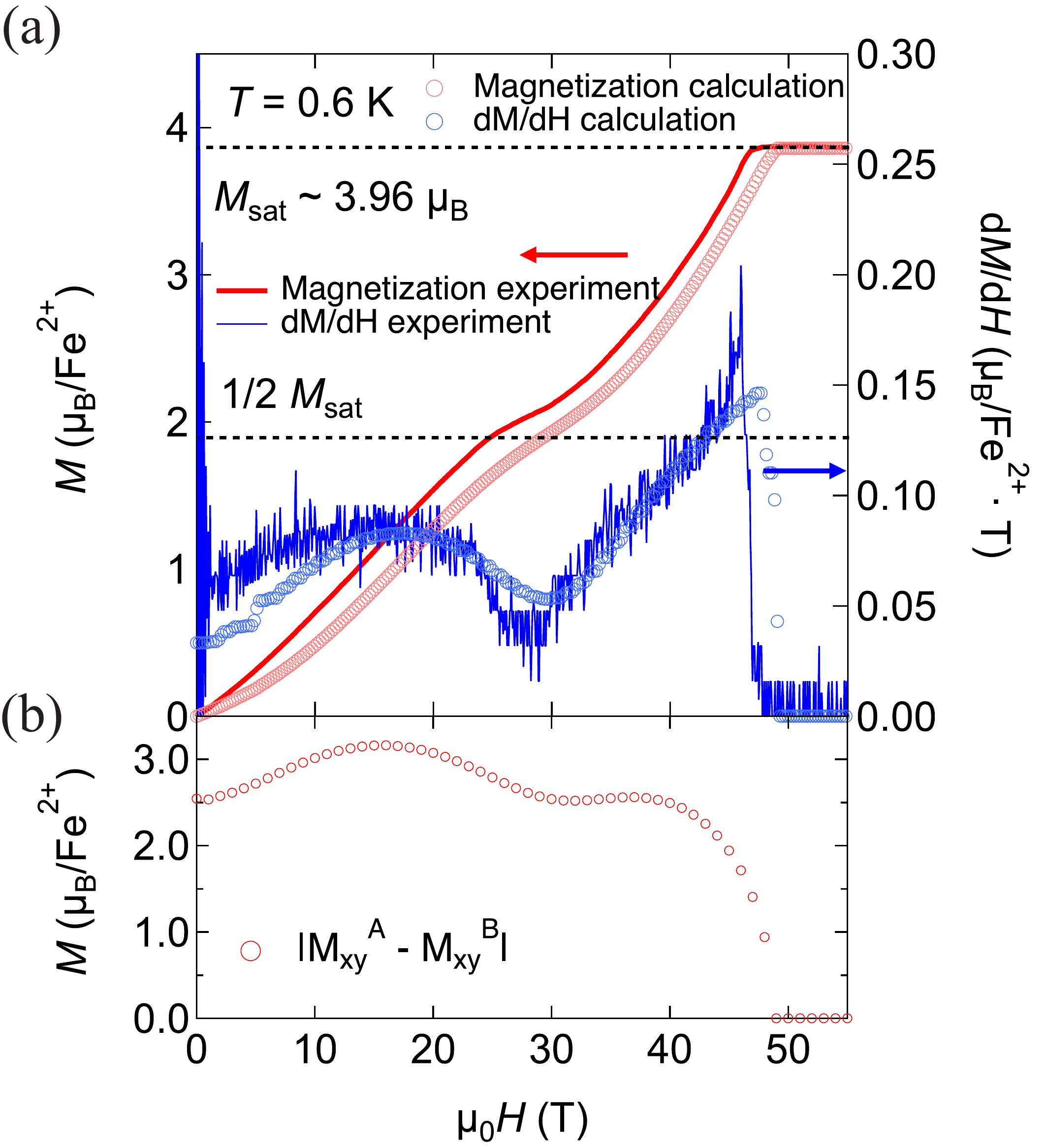}
	\end{center}
	\par
	\caption{\label{fig_magnetization} {\bf Magnetization in \bfso{}}  (a) Data (solid lines) and calculations (open symbols) for the  pulsed field magnetization $M$ vs magnetic field $H$ at  $T = 0.6$ K (red curve). $M(H)$ is shown in red on the left axis and $dM/dH$ in red on the right axis. The calculated magnetization (red open circle) and its derivative with field (red open circle) is the out-of-plane net moment $\left|M_{z}^{A} + M_{z}^{B}\right|$. (b) The calculated staggered moment $\left|M_{xy}^{A} - M_{xy}^{B}\right|$ at zero temperature.}
\end{figure}

We present mean field calculations of $M(H)$ along with the experimental data in \autoref{fig_magnetization}. Experimental $M(H\parallel c$) at $T = 0.6$ K shows a sharp saturation around 48 T with $M_{\text{sat}} \sim$ {\color {black} 3.86} $\mu_{B}$/Fe$^{2+}$. Accordingly, $dM/dH$ plunges to zero above 48 T. The saturation magnetization is consistent with the $\bf{S} = 2$ value and $g_{c} = 1.93$. Interestingly, $M(H)$ shows a slanted plateau around 30 T, where $dM/dH$ displays its broad dip. The magnetization value at the minimum in $dM/dH$ is around 2 $\mu_{B}$/Fe$^{2+}$ close to  $M_{\text{sat}}/2$, consistent with saturation of the $\left|S^z = 1\right>$ state. The end of the plateau is thus likely due to the subsequent crossing of the $\left|S^z = 2\right>$ state with the levels that were the ground state at lower fields. The existence of the half-plateau is in contrast to the magnetization of {\color {black}another Melilite} compound Ba$_{2}$CoGe$_{2}$O$_{7}$ that shows no magnetization plateau until saturation \cite{Kim_multiferroic}. 

\begin{figure}[tb]
	\linespread{1}
	\par
	\begin{center}
		\includegraphics[width=0.5\textwidth]{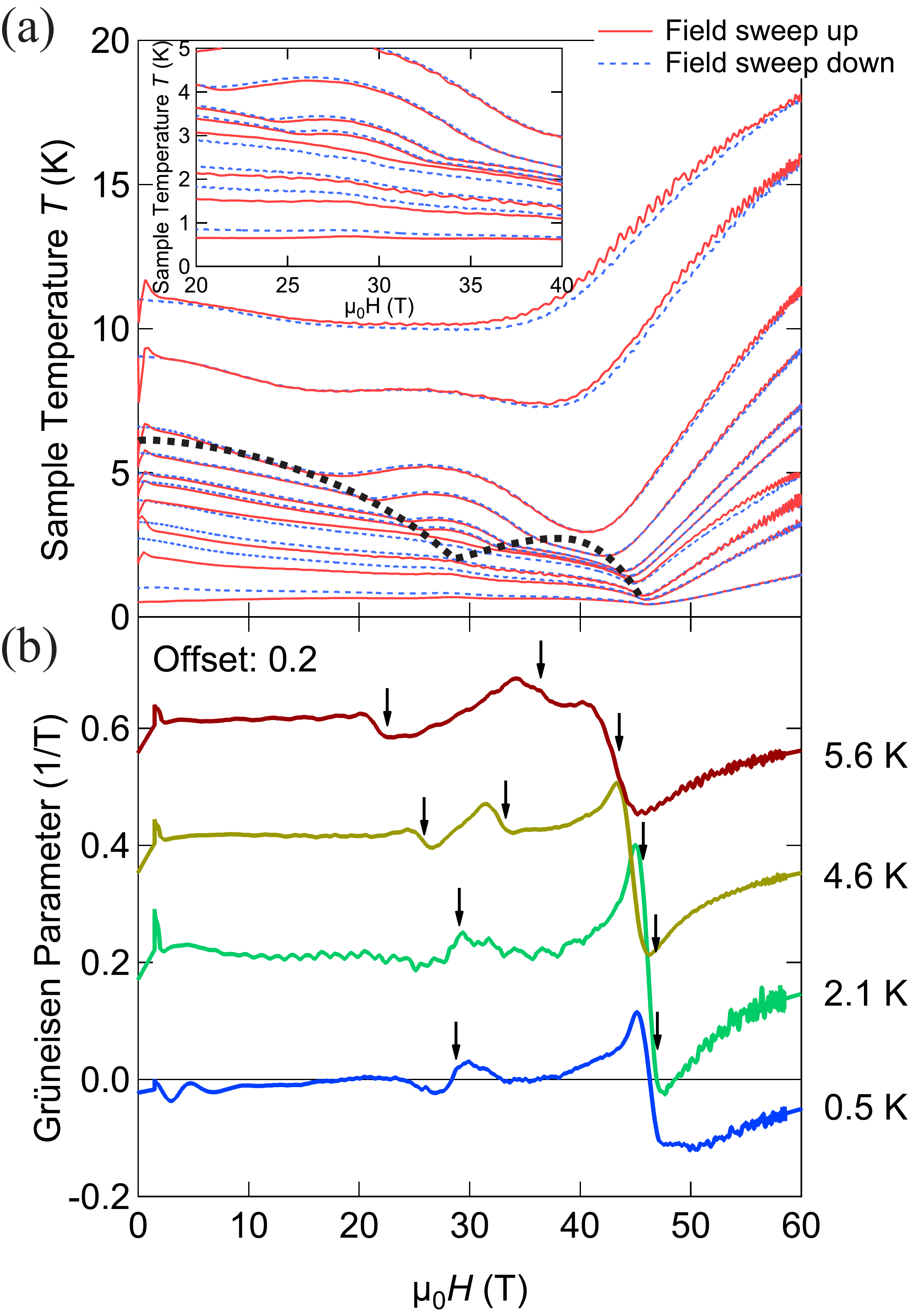}
	\end{center}
	\par
	\caption{\label{fig_magnetocaloric} {\bf Magnetocaloric effect in \bfso{}} (a) Magnetocaloric effect measurements - {\color {black} sample }temperature $T$ vs magnetic field $H$ under quasi-adiabatic conditions in pulsed magnetic fields with several starting temperatures. The red solid lines are field charging and the red dashed lines are field discharging. The thick black dashed line is a guide to eye for the phase boundaries. {\color {black} The inset is an enlargment of the data between 20 and 40 T.} (b) Grüneisen parameter at different temperatures vs magnetic field calculated from the magnetocaloric effect. The black arrows indicate the points where the Grüneisen parameters sign changes occur. Curves are offset.}
\end{figure}
\begin{figure*}[tbp]
	\linespread{1}
	\par
	\begin{center}
		\includegraphics[width=.9\textwidth]{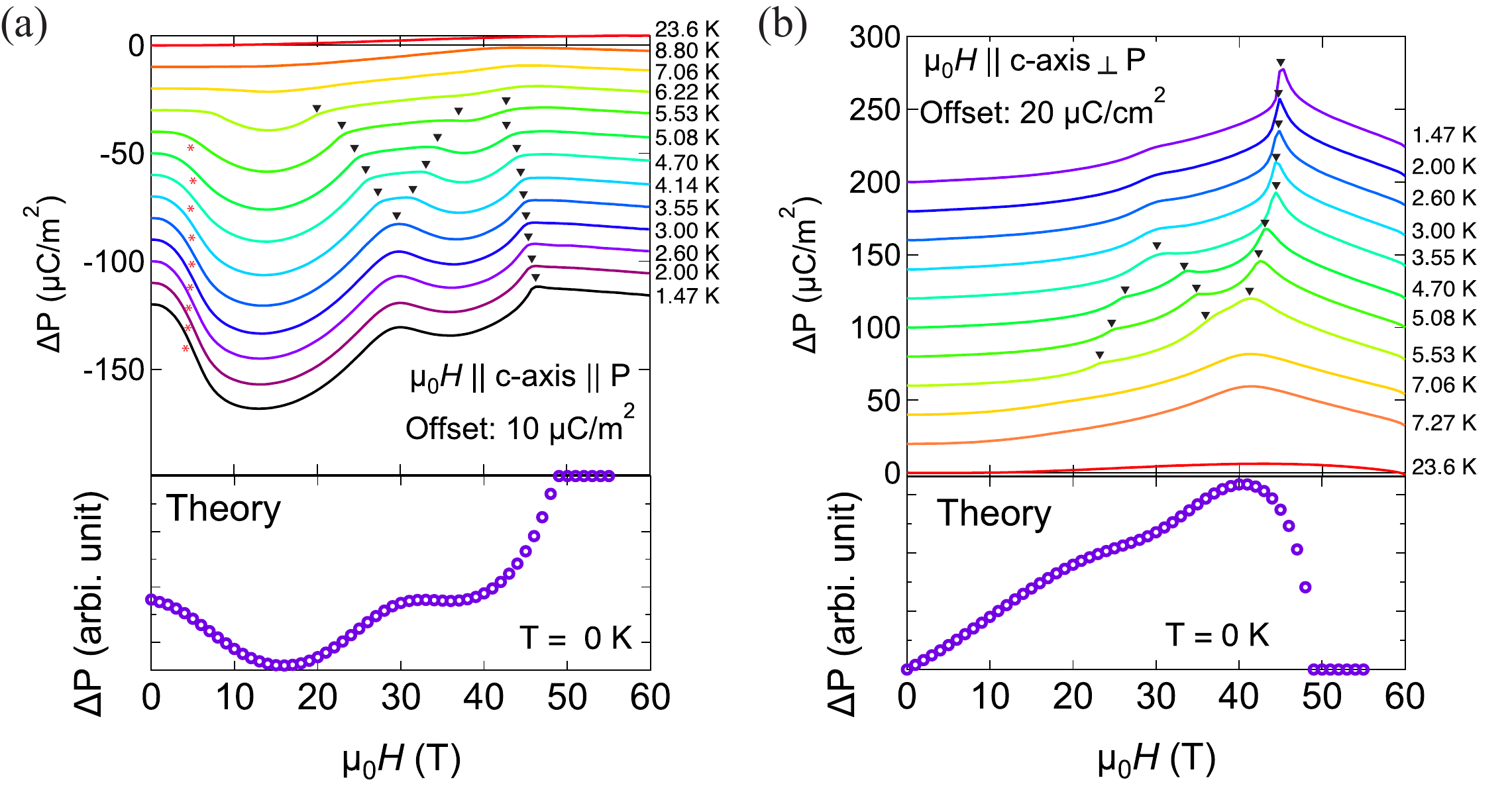}
	\end{center}
	\par
	\caption{\label{fig_polarization} {\bf Ferroelectricity in \bfso{}} Field dependence of the electric polarization relative to zero magnetic field, $\Delta P$, with magnetic field along the c-axis $H \parallel c$ for (a) ${\bf P} \parallel H$ and (b) ${\bf P} \perp H$. Curves at different temperature $T$ are offset as shown. Top figure: measurements, bottom figures: mean field calculations for zero temperature. 
 }
\end{figure*}

Next we turn to the magnetocaloric effect in pulsed magnetic fields, which is the change in temperature with changing magnetic field $T_{\text{sample}}$($H$) under adiabatic or semi-adiabatic conditions. The magnetocaloric effect is a thermodynamic parameter and sensitive to entropy changes in the system. Since the entropy roughly doubles near a spin level crossing, the magnetocaloric effect shows strong effects. 
$T_{\text{sample}}$($H$) is shown in \autoref{fig_magnetocaloric} (a) for up and down sweeps of the pulsed magnetic field for various starting temperatures in zero field. The temperature curves for up and down sweeps are nearly identical above 1.5 K, suggesting adiabatic conditions. Below 1.5 K, a small hysteresis appears indicating semi-adiabatic conditions (the hysteresis does not appear in {\color {black} DC} magnetic-field measurements or measurements with additional exchange gas, indicating it is not intrinsic to the compound). Starting from 11 K in the paramagnetic state, $T_{\text{sample}}$($H$) decreases slightly with increasing field and exhibits a minimum around 10 K and 35 T. Above 35 T, the $T_{\text{sample}}$($H$) increases up to the highest fields. At these high temperatures, none of the features are sharp and the  changes of curvature arise from changes in the energy gaps between spin levels, and from magnetic short-range correlations. 
Starting from 3 K, which is well below $T_{\text{N}}$, $T_{\text{sample}}$($H$) decreases with increasing field and shows a kink around 25 T and a broad maximum around 30 T followed by another kink around 35 T. These fields are consistent with fields of the half-plateau observed in the magnetization measurement. They also correspond to phase transitions from the antiferromagnetic state to a paramagnetic state (25 T) and then back into the long-range ordered state (35 T), e.g. between the two peaks of the dome in $T-H$ space. The increase in $T_{\text{sample}}$($H$) between 25 T and 30 T outside of the antiferromagnetic state occurs {\color {black}due to the reduced magnetic entropy, since the half-plateau state is closely approximated by the single state $\left|S^{z} = 1\right>$}. {\color {black} As shown in the inset of \autoref{fig_magnetocaloric} (a), the kinks gradually merge and become a broad hump when the temperature of the sample is below 2 K.}

Upon further increase of the field, $T_{\text{sample}}$($H$) shows another sharp kink and a minimum which is followed by a monotonic increase of $T_{\text{sample}}$($H$) up to 60 T. This corresponds to the transition from the AFM state to a fully-polarized spin state. Finally in the fully-saturated phase above 50 T, $T_{\text{sample}}$($H$), continues to increase. This is due to the gap opening between the $\left|S^z = 2\right>$ high-field ground state and the other spin states. The opening gap
reduces the spin entropy, forcing the lattice entropy and temperature in turn to increase to conserve overall entropy.
Other curves that start below $T_{\text{N}}$ at zero field also show similar behavior.  

As shown in \autoref{fig_magnetocaloric} (b), we take the derivatives of $T_{\text{sample}}$($H$) with magnetic field to obtain the Grüneisen parameter $G(H) = \frac{1}{T}\left(\frac{dT}{dH}\right)_{S}$. Interestingly, $G(H)$ at the lowest temperature shows a broad feature with a sign change around the half plateau field at 30 T, followed by a sharp feature with a sign change and a near-vertical slope (divergence) at the phase transition to saturation. The lower-field feature splits into two features with increasing temperature reflecting the double-peaked dome in $T-H$ space. The {\color {black} sign-changing diverging behavior in the magnetic $G(H)$ at the 48 T phase transition close to saturation is a strong indication of quantum critical points formed at a spin level crossings \cite{Gruneisen1,Gruneisen2}. Essentially, the spin level crossing that began at 30 T as the dispersed $|S_z = 2>$ state started to cross the $S_z = 0,1$ states is now completed at the 48 T saturation transition, leaving $S_z = 2$ as the sole ground state. This saturation of the spin state is accompanied by a quantum phase transition from the antiferromagnetic state to the field-aligned $|S_z = 2>$ state.} Finally, at the lowest fields below 5 T, sharp features in $T(H)$ are artifacts of the firing of the capacitor bank that drives the pulsed fields. However, a broad feature near 5 T can also be resolved that will also be discussed in the next section on electric polarization.

We also measure the electric polarization $P(H)$ 
in pulsed magnetic fields, for ($P_{\parallel}$) and perpendicular ($P_{\perp}$) to the applied magnetic field $H \parallel c$, shown in \autoref{fig_polarization} (a) and (b). For comparison, we show the calculated electric polarization in a {\it p-d} hybridization  model~\cite{pdhybridization1,pdhybridization2}.  The S.I.~\cite{SI} also shows the raw data ($dP/dt$) {\color {black} and zoomed version of these figures between 20 T and 40 T. The two sharp kinks in $P(H)$ and ($dP/dt$) above and below 30 T observed at high temperature above 3 K gradually approach and form a broad features at low temperatures below 3 K.} 
{\color {black} The sharp features are consistent with the field-induced transitions in and out of long-range order. Broad features at 30 T are consistent with {\color {black} the end of the magnetization plateau and the onset of the $|S_z = 2>$ spin level crossing the ground state}, which is also seen in the magnetization (\autoref{fig_magnetization}) and  magnetocaloric effect (\autoref{fig_magnetocaloric}) measurements. For ($P_{\parallel}$) (\autoref{fig_polarization} (a)), there is a sharp drop below 5 T consistent with another phase transition, followed by a broad minimum, then a broad {\color {black} hump}
near 30 T consistent with the onset of the spin level crossing, and finally a sharp shoulder at phase transition to magnetic saturation near 50 T. This sharp drop at 5 T also corresponds to a broad peak in the magnetocaloric effect.}
On the other hand, for $P_{\perp}$  in \autoref{fig_polarization} (b) no feature is seen at 5 T, but a series of peaks appear in higher fields. At the lowest temperature there is one {\color {black} broad slope change}, resembling a broad step, near 30 T {\color {black} consistent with the onset of the $|S_z = 2>$ level crossing} followed by a sharp peak at saturation at 48 T. With increasing temperature above 4.7 K, the peak near 30 T splits into two peaks that separate in field (reflecting the edges of the two antiferromagnetic domes in $T-H$ phase space) and finally all features broaden and are eventually suppressed with increasing temperatures. The fact that $\Delta P(H)$ is not completely field-independent above magnetic saturation is likely a result of a slightly changing temperature above 50 T as shown in \autoref{fig_magnetocaloric}. Although the electric polarization and magnetization data are taken in helium exchange gas in an attempt to thermalize the sample, the extraordinarily large magnetocaloric effect in this sample ensures that we do not achieve perfect thermalization in millisecond pulsed fields.

\section{Theory}
\begin{table}[]
\caption{\label{parameter_table}Parameter set of the model Hamiltonian Eq. (\ref{Hamiltonian}) to simulate the magnetization and electric polarization}
{\begin{tabular}{|c|c|c|c|c|c|}
\hline
$J$      & $J'$    & $D$    & $A$   & $C$     & $g_{c}$\\ \hline\hline
0.0887 \text{meV}& {0.00887 \text{meV}}  & 2.47 \text{meV}& 0.1 \text{meV} & -0.15 \text{meV} & 1.93\\ \hline
\end{tabular}%
}
\end{table}
To understand the experimental results, we build a Hamiltonian and compute the magnetization and electric polarization using a classical theory of SU($N$) coherent states~\cite{HZhang_GMFT,Dahlbom22,Dahlbom22b}. In our theoretical model, we retain all $N = 5$ levels for each ${\bf S} = 2$ spin. A system configuration consists of a direct 
product over the lattice sites of coherent states of the SU(5) group of unitary transformations with determinant equal to one
that acts on the local Hilbert space of each spin. Since the Hamiltonian of \bfso{} is not frustrated (the spin lattice is bipartite and the dominant exchange interaction occurs between nearest-neighbor atoms along the in-plane and $c$-directions), we only consider a variational space of two-sublattice mean field states:
\begin{equation}
| \Psi_{\rm MF} \rangle = \otimes_{i \in A} | \Psi_{\rm A} \rangle_i   \otimes_{j \in B} | \Psi_{\rm B} \rangle_j 
\end{equation}
where $| \Psi_{\rm A} \rangle$ and $| \Psi_{\rm B} \rangle$ are normalized SU(5) coherent states on the $A$ and $B$ sublattices, respectively,
\begin{eqnarray}
|\Psi_{\rm A} \rangle = \!\!\! \sum_{j=-2,2}  z^A_j | S^z = j \rangle,
\;\;\;
|\Psi_{\rm B} \rangle = \!\!\! \sum_{j=-2,2}  z^B_j | S^z = j \rangle,
\end{eqnarray}
with $\sum_j | z^A_j|^2 = \sum_j | z^B_j|^2=1$. The mean field state $| \Psi_{\rm MF} \rangle$ is obtained by minimizing the classical energy $E_{\rm MF} = \langle \Psi_{\rm MF}| \mathcal{H}  | \Psi_{\rm MF} \rangle $ as a function of the variational coefficients $z^A_j$ and $z^B_j$.

As mentioned above, the combination of the crystal field of the tetragonally compressed FeO$_{4}$ and the relatively strong spin-orbit coupling ($\sim$ 20 meV) induces a predominant easy-plane anisotropy ($D$) along $c$-axis (see \autoref{fig_crystalstructure} (d) \cite{THJang_PRB,intro7,SFSO}). Moreover, the large spin of Fe$^{2+}$ also allows for small single-ion anisotropy terms that are quartic in the spin components. We do not consider the Dzyaloshinskii-Moriya (DM) interaction because in contrast to the case of  Ba$_{2}$CoGe$_{2}$O$_{7}$ \cite{BCGO_multiferroic1},  weak ferromagnetism is absent in \bfso{}.
We are also neglecting interactions between neighboring quadrupole moments~\cite{Romhanyi11,Soda14}, {\color {black} because} they are not expected to produce a significant change in the field dependence of the magnetization and the electric polarization. 

We then use the following minimal Hamiltonian: 
\begin{eqnarray}
\nonumber \mathcal{H} &=& J\sum_{i,j} {\bf S}_i \cdot {\bf S}_j + J'\sum_{\langle\langle i,j\rangle\rangle} {\bf S}_i \cdot {\bf S}_j
\\
\nonumber&+&D\sum_{i}\left(S_{i}^{z}\right)^{2} - h\sum_{i}S_{i}^{z} 
\nonumber \\
&+&  A\sum_{i \in A}\left[\left( {\bf S}_{i} \cdot {\bf n}^A_1 \right)^{4} + \left({\bf S}_i \cdot {\bf n}^A_2 \right)^{4}\right] 
\nonumber \\
&+& 
A\sum_{j \in B}\left[\left( {\bf S}_{j} \cdot {\bf n}^B_1 \right)^{4} + \left({\bf S}_j \cdot {\bf n}^B_2 \right)^{4}\right] 
\nonumber \\
&+&  C\sum_{i}\left(S_{i}^{z}\right)^{4},
\label{Hamiltonian}
\end{eqnarray}
where the first and second terms are the nearest-neighbor intra- and interplane interactions of the square lattice. The third term is  the easy-plane  single-ion anisotropy ($D > 0$), and the fourth term is a Zeeman energy term where $h$ is the effective external magnetic field that absorbs the product of the $g_{c}$ factor and the Bohr magneton: $h = g_c \mu_B H$. $A$ and $C$ are {\color {black} small} quartic single-ion anisotropy terms. We note that the $A$-term is different for both sublattices because of the staggered rotation along the $c$-axis of the 
FeO$_4$ tetrahedra. The in plane unit vectors $\{ {\bf n}^A_1, {\bf n}^A_2\} $ and $\{ {\bf n}^B_1, {\bf n}^B_2\} $ correspond to the in-plane orthogonal directions of the principal axes of the FeO$_4$ tetrahedra in the A and B sublattices, respectively. 
The $x$, $y$, and $z$ spin components are parallel to  the crystallographic $[110]$, $[1\bar{1}0]$,  and $c$ axes, respectively, as shown in \autoref{fig_crystalstructure} (a).

Among these spin energy terms, the easy-plane anisotropy term $D(S^z)^2$ is dominant, hence the description of the magnetic properties begins with a single ion picture: the five states of Fe$^{2+}$ split into three groups of one singlet with $\left|S^{z} = 0\right>$ at zero energy, one doublet of $\left|S^{z} = 1\right>$ {\color {black} an energy $D$ above the ground state} and another doublet of $\left|S^{z} = 2\right>$ at $4D$ {\color {black} above the ground state}. The $J$ and $J'$ terms create finite bandwidth in the levels. On account of the strong single-ion anisotropy, we employ the above-mentioned mean field theory at {\it T} = 0 K \cite{HZhang_GMFT,Dahlbom22,Dahlbom22b} to calculate the field-induced magnetization  component along the $c$-axis,   $M= g_c \mu_B \langle \Psi_{\rm A}| S^z_{j \in A} |  \Psi_{\rm A} \rangle =  g_c \mu_B \langle \Psi_{\rm B}| S^z_{j \in B} |  \Psi_{\rm B} \rangle$ and $dM/dH$.

Our mean-field calculations 
 reproduce qualitatively and quantitatively both the experimentally measured $M(H)$ and $dM/dH$ very well as shown in \autoref{fig_magnetization} (a). \autoref{fig_magnetization} (b) shows the calculated staggered magnetic moment and total magnetic moment within the $ab$-plane. The staggered magnetic moment shows broad near 15 and 40 T and shows a minimum corresponding to the plateau in $M(H)$.  The in-plane total magnetic moment remains zero up to 55 T. Therefore, the in-plane magnetic moments of the two sublattices maintains their collinear spin structure at zero in the whole field range. On the other hand, the total out-of-plane magnetic moment of two sublattice increases monotonically with magnetic field as shown in \autoref{fig_magnetization} ({\color {black} a}), creating an overall canted spin structure. The staggered moment of out-of-plane component of spins remains zero. Our experiments measure the total out-of-plane magnetization, thus our calculation 
 reproduces the experimental data, which support that the spin Hamiltonian we used is reliable. More importantly, it also confirms that the system has an extremely strong uniaxial easy-plane magnetic anisotropy in comparison with the exchange parameters, {\color {black} $(D/J)$ $>$ 27.}
 We summarize the parameters employed in this calculation in \autoref{parameter_table}. Our values match well with the parameter obtained from inelastic neutron scattering study \cite{intro7}.

We also calculate the electric polarization vs magnetic field, which can be expected from the fact that the lattice in combination with the magnetic order breaks symmetry to create a polar axis, making \bfso{} a type II multiferroic.
In most multiferroics, there are two categories of microscopic mechanisms by which magnetism can create an electric polarization: (1) the rearrangement of electron density in orbitals, and (2) distortion of the lattice and the relative location of the charged ions \cite{Eerenstein07,Khomskii09,Cheong07}.
These rearrangements occur in order to lower magnetic energy, at the expense of the energy cost of rearranging the lattice. From symmetry arguments, both categories of microscopic mechanisms should produce qualitatively similar electric polarization vs magnetic field \cite{Perez-Mato11}. Our attempt to measure magnetostriction in this compound (data not shown) to a sensitivity of 1 part in 10$^{7}$ showed  no resolvable magnetostriction, similar to another silicate easy-plane magnet BaCu$_2$SiO$_6$.\cite{Marcelo} Thus the electronic orbital rearrangement might be expected to be the dominant mechanism and we neglect magnetostriction effects. 
Here we use the microscopic {\it p-d} hybridization model~\cite{pdhybridization1,pdhybridization2} (which models the rearrangement of electron density in orbitals) to calculate the electric polarization as was done for other Melilite compounds \cite{BCGO_multiferroic1}. Models based on magnetostriction show qualitatively similar $P(H)$ contributions in Ba$_2$CoGe$_2$O$_7$ \cite{BCGO_multiferroic2,Perez-Mato11} and in the canted {\it XY} antiferromagnet NiCl$_2$-4SC(NH$_2$)$_2$ \cite{Zapf11} with a polar lattice - generally different multiferroic electric polarizations that derive from the same symmetry breaking effects will have qualitatively similar field and temperature depedencies, regardless of the microscopic mechanisms. 
We use an orbital hybridization scheme \cite{pdhybridization1,pdhybridization2} to find the electric polarization of \bfso{} based on the spin structure from the spin Hamiltonian Eq. (\ref{Hamiltonian}) using the following formula \cite{Romhanyi11}:
\begin{eqnarray}
\hat{P}_{j}^{x} \!\!\! &\propto& \!\!\! -\cos 2 \kappa_{j}\left(\hat{S}_{j}^{x} \hat{S}_{j}^{z}+\hat{S}_{j}^{z} \hat{S}_{j}^{x}\right)-\sin 2 \kappa_{j}\left(\hat{S}_{j}^{y} \hat{S}_{j}^{z}+\hat{S}_{j}^{z} \hat{S}_{j}^{y}\right), 
\nonumber \\
\hat{P}_{j}^{y} \!\!\! &\propto& \!\!\! \cos 2 \kappa_{j}\left(\hat{S}_{j}^{y} \hat{S}_{j}^{z}+\hat{S}_{j}^{z} \hat{S}_{j}^{y}\right)-\sin 2 \kappa_{j}\left(\hat{S}_{j}^{x} \hat{S}_{j}^{z}+\hat{S}_{j}^{z} \hat{S}_{j}^{x}\right), 
\nonumber \\
\hat{P}_{j}^{z} \!\!\! &\propto& \!\!\! \cos 2 \kappa_{j}\left[\left(\hat{S}_{j}^{y}\right)^{2}-\left(\hat{S}_{j}^{x}\right)^{2}\right]-\sin 2 \kappa_{j}\left(\hat{S}_{j}^{x} \hat{S}_{j}^{y}+\hat{S}_{j}^{y} \hat{S}_{j}^{x}\right)
\nonumber \\
\label{Eq:pol}
\end{eqnarray}
where $j$ belongs to either sublattice $A$ or $B$. The different orientation of the tetrahedra is accounted for by choosing $\kappa_{j \in A}=\kappa$ and $\kappa_{j \in B}=-\kappa$, where $\kappa \simeq 21^{\circ}$ is the rotation angle along the $z$-axis. Note that the the three components of the polarization operators are proportional to linear combinations of the  quadrupole moments 
$\hat{O}_{x z}= \hat{S}^{x} \hat{S}^{z}+\hat{S}^{y} \hat{S}^{z}$, $\hat{O}_{y z}=\hat{S}^{y} \hat{S}^{z}+\hat{S}^{z} \hat{S}^{y}$,
 $\hat{O}_{x y}=\hat{S}^{x} \hat{S}^{y}+\hat{S}^{y} \hat{S}^{x}$, and $\hat{O}_{2}^{2}=\left(\hat{S}^{x}\right)^{2}-\left(\hat{S}^{y}\right)^{2}$.
 In other words, the dynamic response of the electric polarization can be extracted from the dynamic quadrupole susceptibility, which is directly obtained from a generalized SU(5) spin wave theory or, equivalently, from a linearization of the classical dynamics based on coherent states of SU(5)~\cite{HZhang_GMFT,Dahlbom22,Dahlbom22b}.

The mean field expectation value of the net electric polarization per site, 
$ \langle \Psi_{\rm MF} | ( \frac{1}{N} \sum_j {\bf P}_j) | \Psi_{\rm MF} \rangle$,  matches nearly all of the experimentally-observed features reasonably well. The results of calculated $P(H)$ for ${\bf P} \parallel c$ and ${\bf P} \perp c$ are presented in \autoref{fig_polarization}. $P(H)$ for ${\bf P} \parallel c$ matches the experimental data extremely well, while for ${\bf P} \perp c$ the overall shape matches but the theoretical predictions show a broader peak near the 47 T saturation field than the experiments. One notable discrepancy between simulation and measurement appears above the saturation field for ${\bf P} \perp H$. Here a temperature instability in the experiments due to the magnetocaloric effect likely drives the electric polarization to evolve above magnetic saturation. One other small discrepancy between experiment and theory is the feature near 5 T for ${\bf P} \parallel H$. Here both experiment and theory show a downturn with an inflection, however in experiments this feature is more pronounced. The magnetocaloric effect in  \autoref{fig_magnetocaloric} also shows a sharp feature at this field, and may indicate an additional phase transition whose origin is not clear in this study.
Overall the theory captures the major features of the electric polarization and the magnetization.

\section{discussion}
\begin{figure}[tbp]
	\linespread{1}
	\par
	\begin{center}
		\includegraphics[width=0.5\textwidth]{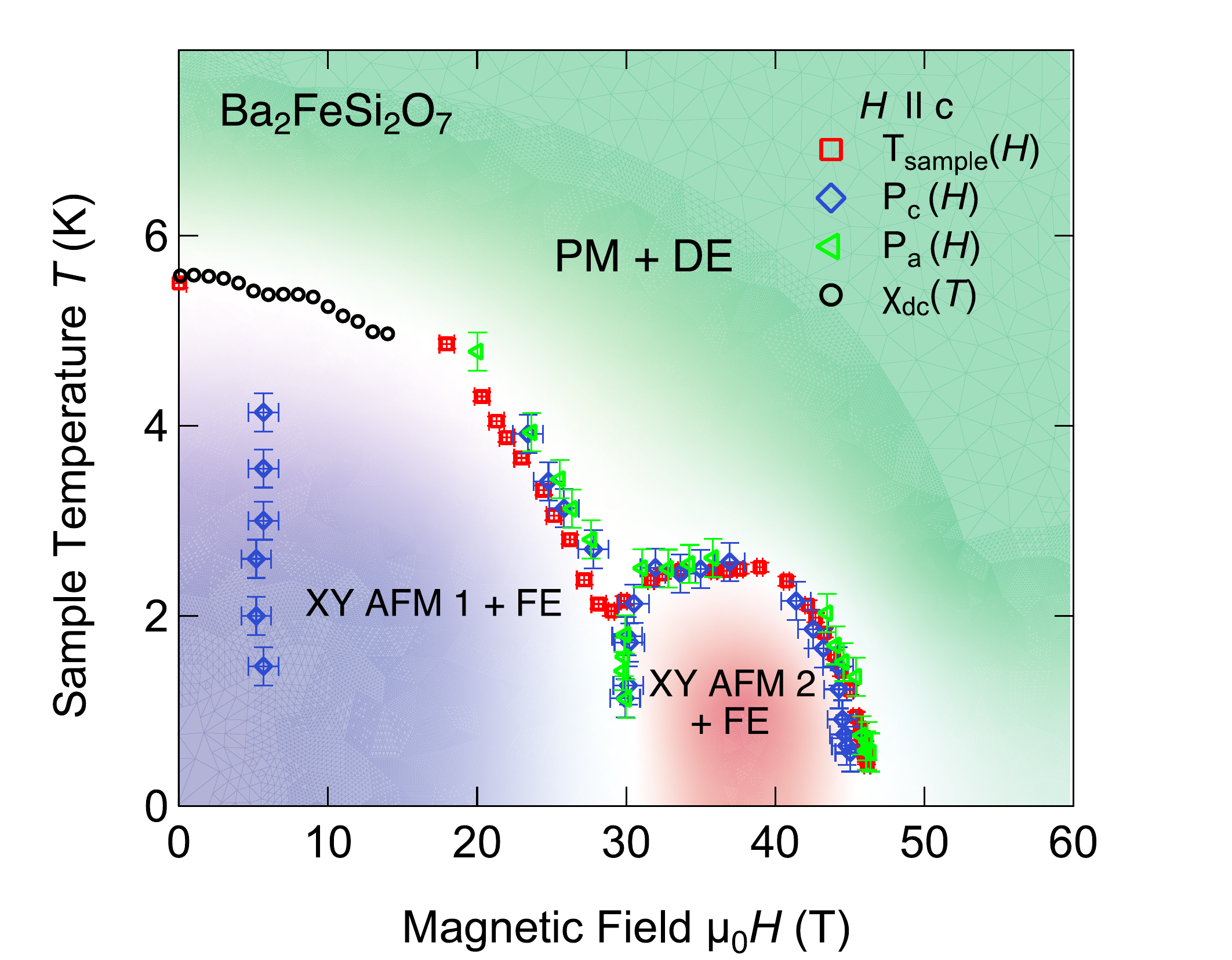}
	\end{center}
	\par
	\caption{\label{fig_phasediagram} {\bf Magnetic phase diagram of \bfso{}}  Experimental temperature-magnetic field phase diagram of Ba$_2$FeSi$_2$O$_7$ based on {\color {black} DC} magnetic susceptibility $\chi_{dc}$, electric polarization and magnetocaloric effect measurements in pulsed magnetic fields. Regions of quasi-{\it XY} antiferromagnetism ({\it XY} AFM), ferroelectricity (FE), paramagnetism (PM) and dielectric behavior (DE) are indicated. There is no phase transitions observed between the two domes.}
\end{figure}
An experimental $T-H$ phase diagram was constructed by combining the data from \autoref{results} as shown in \autoref{fig_phasediagram}. 
We observed two dome-shaped magnetic phases in addition to the paramagnetic phase. These two domes overlap each other at 30 T where we observed the end of a slanted half-plateau in $M_z(H \parallel c)$, {\color {black} a weak hump in $T(H)$ in the magnetocaloric effect, and broad slope change in polarization}. We denote the first dome and second dome as AFM1 and AFM2, respectively. Let us first discuss the AFM1 phase. According to our mean-field calculations and previous neutron diffraction data, the spins at zero field are confined to the {\it ab} plane, forming an {\it XY} AFM phase. With increasing magnetic field, $S^{z}$ increases while maintaining {\it XY} AFM in the plane. 

In this AFM1 state, the spins are mostly a superposition of the overlapping $\left|S^z = 0\right>$ state and the $\left|S^z = 1\right>$ band. 
Thus, an effective low-energy ${S} = 1$ model and the corresponding  SU(3) spin wave theory worked well to explain the inelastic neutron scattering for the AFM1 phase \cite{intro7,SU3_2}. We also find field-induced changes in electric polarization consistent with multiferroic behavior (\autoref{fig_polarization}). Although the crystal symmetry without magnetic order is not polar, when the {\it XY} spin ordering is combined with it, the overall system acquires a polar axis that points along the $c$-axis at zero field and tilts towards the {\it ab} plane with increasing field. This is consistent with our measurements. 

In the AFM2 phase, with increasing field, the dispersed $\left|S_{z} 2\right>$ band starts to cross and become degenerate with the ground state. The AFM2 state is described by $\psi_{02} \text{ } {\color {black} \sim} \left|S^{z} = 1\right> + \left|S^{z} = 2\right>$ between 30 T and 55 T. As the mean field calculation demonstrates, the AFM2 state maintains an {\it XY} AFM in-plane spin component while the out-of-plane component monotonically increases. Theoretically there are no symmetry changes between AFM1 and AFM2 and experimentally there are no indications of a phase transition between them {\color {black} as demonstrated in magnetization, magnetocaloric effect and polarization measurements}. Thus we conclude AFM1 and AFM2 are qualitatively same phase and the half-plateau is a spin level {\color {black} crossing}. This is in contrast to Ba$_3$Mn$_2$O$_8$ that shows two separate magnetic order domes \cite{Samulon09}. Since AFM1 and AFM2 have the identical symmetry, the same symmetry analysis for the multiferroicity of AFM1 phase can be applied for AFM2 phase. Thus, the gradual increase in electric polarization perpendicular to magnetic field and the upward concave in polarization parallel with magnetic field are the common feature in both AFM1 and AFM2. 

\section{CONCLUSION}
To summarize, we observe a canted {\it XY} antiferromagnetic phase in which the spin state changes due to a field-induced spin level crossing, but preserves the symmetry of the magnetic ordering. The result is a double-dome feature in the temperature-field phase diagram up to 60 T. This phenomenology is made possible by the strong uniaxial anisotropy term in \bfso{} that splits different spin levels, allowing them to be tuned by applied magnetic fields. AFM order occurs in the {\it ab} plane with canting out of the field in increasing field. The two domes in $T-H$ space (AFM1 and AFM2) correspond to two different spin states of the ${S} = 2$ Fe$^{2+}$ spins, e.g. a superposition of mostly $\left|S^z = 0\right>$ and $\left|S^z = 1\right>$ for AFM1 and mostly $\left|S^z = 1\right>$ and $\left|S^z = 2\right>$ for AFM2. {\color {black} We note that the spin levels should also be thought of as dispersed bands due to exchange coupling, and thus their crossings occur over extended field ranges, resulting in the finite width of the AFM domes in magnetic field.} We demonstrated that no symmetry breaking occurs between two domes and they are smoothly connected as both AFM1 and AFM2 form antiferromagnetic {\it XY} long-range ordering with spin canted along {\it c}-axis. Between these two domes, the magnetization vs magnetic field along the {\it c}-axis shows a slanted half-plateau {\color {black} corresponding to the $|S_z = 1>$ state becoming the ground state while the $|S_z = 0>$ state is pushed to higher energy. The plateau ends when the $|S_z = 2>$ state begins to cross the ground state.} We also measure and calculate the electric polarization due to type II multiferroic behavior as the symmetry-breaking of the magnetism in conjunction with the lattice creates a polar axis. We build a spin Hamiltonian that includes in-plane and out-of-plane Heisenberg exchange interactions, as well as {\color {black} small} quadratic and quartic  single-ion anisotropy terms, which explains well the experimental data qualitatively and quantitatively. We employ a  mean field approach based on SU(5) coherent states, that is necessary to capture all five $S^z$ levels of the ${S}= 2$ spin of Fe$^{2+}$. Intriguing optical phenomena such as directional dichroism and magnetochiral effects in the multiferroic phase will be also very interesting future study on this material. Such an ordered phase containing a spin level/band crossing in the middle of it is unusual and possibly unique.  It results from the interplay between uniaxial anisotropy that creates well-separated spin levels, and magnetic exchange that causes those levels to mix and form long-range magnetic order in certain regions of phase space.

\begin{acknowledgments}
 Scientific work at LANL was primarily funded by the LDRD program at LANL. The facilities of the NHMFL are funded by the U.S. NSF through Cooperative Grant No. DMR-1157490, the U.S. DOE and the State of Florida. Theoretical work at UTK by D.D. and C.D.B. was funded by the U.S. Department of Energy, Office of Science, Office of Basic Energy Sciences, under award DE-SC-0018660. R. S. acknowledges support by the G. T. Seaborg Institute Postdoctoral Fellow Program under project No. 20210527CR. S. D. and A. C. at ORNL were supported by the U.S. Department of Energy, Office of Science, Basic Energy Sciences, Materials Sciences and Engineering Division. M. J. (experimental contribution) acknowledges support by DOE Office of Science B.E.S project "Science at 100 Tesla". The work at MPK/POSTECH (sample growth) was supported by Grant No. 2020M3H4A2084418 and Grant No. 2022M3H4A1A04074153, through the National Research Foundation (NRF) funded by MISP of Korea.
\end{acknowledgments}


\begin{thebibliography}{50}
\bibitem{Marcelo} M. Jaime, Los Alamos National Lab, personal communication that multiple attempts to measure magnetostriction in BaCuSi$_2$O$_6$ to 1 part in 10$^{7}$ found no resolvable magnetostriction.
\end{thebibliography}
\end{document}